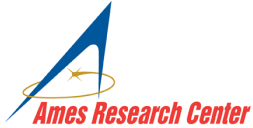 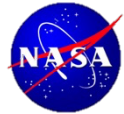

# *THE VALUE OF ASTROMETRY FOR EXOPLANET SCIENCE*

National Academy of Sciences
The Committee on Exoplanet Science Strategy call for white papers

## WHITE PAPER

Eduardo A. Bendek[1,2], Mark S. Marley[1], Michael Shao[3], Olivier Guyon [4], Ruslan Belikov[1], Peter Tuthill[5]

[1]*NASA Ames Research Center, Moffett Field, CA, 94035*
[2]*Bay Area Environmental Research Institute, Moffett Field, CA, 94035*
[3]*Jet Propulsion Laboratory, Pasadena, CA, 91109*
[4]*University of Arizona, Tucson, Arizona 85721, USA*
[5]*Sydney Institute for Astronomy, University of Sydney, NSW 2006, Australia*

**Submitting author information**
*Dr. Eduardo Bendek*
*NASA Ames Research Center / BAERI*
*Email: eduardo.a.bendek@nasa.gov*
*Ph:+1-650-604-5850*

*March 9, 2018*

*Abstract*

Exoplanets mass measurements will be a critical next step to assess the habitability of Earth-like planets: a key aspect of the 2020 vision in the previous decadal survey and also central to NASA's strategic priorities. Precision astrometry delivers measurement of exoplanet masses, allowing discrimination of rocky planets from water worlds and enabling much better modeling of their atmosphere improving species retrieval from spectroscopy. The scientific potential of astrometry will be enormous. The intrinsic astrophysical noise floor set by star spots and stellar surface activity is about a factor of ten more benign for astrometry than for the more established technique of Radial Velocity, widening the discovery region and pushing detection thresholds to lower masses than previously possible. On the instrumental side, precision astrometry is limited by optical field distortion and detector calibration issues. Both technical challenges are now being addressed successfully in the laboratory. However, we have identified the need to continue these technology development efforts to achieve sub-microarcsecond astrometry precision necessary for detection and characterization of Earth-like planets around nearby FGK stars. The international community has realized the importance of astrometry, and various astrometry missions have been proposed and under development, with a few high profile missions now operational. We believe that it is vital for the U.S. scientific community to participate in the development of these new technologies and scientific discoveries. We recommend exploring alternatives to incorporate astrometric capabilities into future exoplanet flagship missions such as HABEX and LUVOIR, substantially increasing the scientific return associated with the expected yield of earth-like planets to be recovered. This white paper contains most conclusions of SAG-12 study group devoted to astrometry.

## 1. Importance of masses for exoplanet science

Several strategic scientific documents such as the 2010 New Worlds New Horizons (NWNH) decadal survey and the NASA Science Plan discuss the need to "Discover and study planets around other stars and explore whether they could harbor life." Such goals are challenging, and we still have not developed the technology necessary to achieve them. The study of terrestrial planets is specifically mentioned on the NWNH 2020 vision chapter which states the importance of finding and studying them around nearby stars, "search for nearby, habitable, rocky or terrestrial planets with liquid water and oxygen… ". Responding to these scientific questions poses a daunting challenge to the scientific and engineering communities.

Mass plays an essential role in the evolution of a planet's interior and atmosphere, whether it is terrestrial or giant. Atmospheric dynamics, outgassing, escape, photochemistry, and a host of other processes directly or indirectly connect back to mass. Thus, any understanding of the context of a planet and how it fits into its dynamical and chemical environment require a mass measurement. Since mass is exceptionally difficult to constrain from only spectral data, the crucial context for comparative planetary science is missing without independent mass constraints. One example from the transiting terrestrial planets is the apparent transition from rocky to gaseous worlds that appears to occur around a few Earth masses (Grasset et al. 2009). Without mass constraints this important demarcation line and the implications it raises for how planets accrete and retain atmospheres remains unknown (Zahnle 2016).

Also, mass measurements offer a unique synergy with spectroscopic observations and direct imaging of extrasolar planets to study and characterize their atmospheres. Discerning the atmospheric composition requires interpretation of reflected light spectra which are strongly sculpted by clouds, hazes, gaseous absorption, and Rayleigh scattering. Unlike the case for transiting planets with measured RV amplitudes, directly imaged planets do not have strong constraints on either their masses or radii. This means that atmospheric scale height, which is inversely proportional to gravity, is an additional uncertainty which must be solved for in retrieval analyses of planetary spectra. Also, since in general the orbital phase at which a planet is imaged will not be perfectly constrained, the uncertain star-planet-observer scattering angle must also be allowed to vary. These two additional uncertainties of unknown gravity and orbital phase substantially expand the range of models which fit a given observation. This means that any inference made on the abundance of atmospheric absorbers is far more uncertain than if there were independent constraints on these quantities (Nayak et al. 2017).

Astrometry also allows expansion of the discovery space for long-period planets around nearby stars of all spectral types, particularly Earth-sized planets in the HZ and beyond the ice line around FGK stars. Furthermore, astrometry can constrain planetary system inclination to determine the mass of RV detected planets and determine whether multiple systems are coplanar.

So far astrometry is the only technique that can unequivocally measure exo-planet masses regardless of the system alignment and it serves to independently confirm direct imaging detections and more precisely measure planet orbits than direct imaging alone. This is particularly important for the case of future flagship missions that expect to yield only a handful of Earth-like planets, hence the likelihood of a transiting geometry is very low preventing absolute mass measurements with RV or Transit Timing Variation technique.

## 2. Imaging stellar astrometry versus RV

Current state-of-the-art instrumentation for exoplanet detection is mostly based on transit photometry and radial velocity (RV) measurements. However, these techniques cannot measure exoplanet masses by themselves and are intrinsically more sensitive to short period planets around dwarf stars. This is especially important in the quest for detecting and characterizing earth-like planets that could populate the Habitable Zone (HZ) of sun-like stars. Planets within the HZ exhibit longer periods and larger angular separations compared to hot planets commonly detected with RV and transit methods. For these reasons, planets located in the HZ and beyond the ice line, are dimmer and exhibit a larger astrometry signal than their hotter counterparts. This regime of higher angular separation, which can be resolved with smaller telescopes, and high contrast corresponds to the operational regime of space telescopes, as shown in Fig.1. In contrast, planets that exhibit a large RV signals have small angular separation and low contrast making them ideal for observation with large telescopes from the ground. Moreover, stellar astrometry probes planetary regions that are difficult to reach using RV and transit measurements. RV's sensitivity reduction to planets with larger Semi-Major Axes (SMA) limits its ability to detect habitable planets.

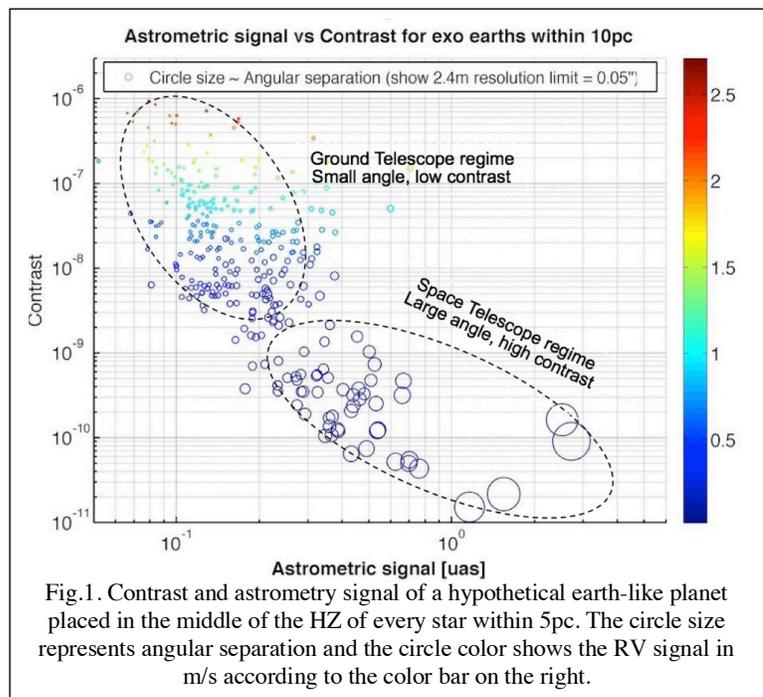

Fig.1. Contrast and astrometry signal of a hypothetical earth-like planet placed in the middle of the HZ of every star within 5pc. The circle size represents angular separation and the circle color shows the RV signal in m/s according to the color bar on the right.

## 3. Astrometry precision

Unresolved brightness inhomogeneity on the star's surface such as star spots can bias the stellar position. The scientific community has studied this problem and has concluded that a typical sun-like star at 10pc viewed from equator would exhibit 0.087µas jitter (Marakov et al. 2009). Other studies showed consistent results of 0.07µas jitter (Lagrange et al. 2011). Thus, astrophysical noise does not impose a fundamental limitation to detect and measure masses of Earth-like planets in the Habitable Zone of sun-like stars at 10pc, where the signal is in the order of 0.3µas. Astrometry holds great promise, not only because the scientific value of masses knowledge, but also because the **Astrophysical limitations** are less critical that in other techniques. if the presence of spots due to stellar activity is the ultimate limiting factor for planet detection, the mass sensitivity of astrometry measurements for Earth-like planets in habitable zones is about *an order of magnitude better* than the sensitivity of prospective ultra-precise RV observations of nearby stars (Makarov et al. 2009).

**Instrumentation limitations**. Although astrometric precision recovered from sparse stellar imaging is ultimately limited by photon noise, errors introduced by non-systematic stellar PSF imaging registration pose a serious challenge before this limit is approached. These are caused by two main sources: time varying optical distortion and detector response changes. Non-systematic *dynamic distortions* that arise from perturbations in the optical train (Benedict et al. 1994; Guyon et al. 2012a; Trippe et al. 2010) will bias the position of background objects that are used to measure the motion of the target star. Distortion grows non-linearly with field of view (FoV), aggravating this problem rapidly as FoV increases. Larger FoV allows imaging more background stars, hence averaging down noise due to peculiar velocities and perturbations induced by unknown binary reference stars. Even in the most stable space environment optics suffers from dynamic distortions in the optical system at the $\mu$as level. For example, Hubble, has been able to achieve 25$\mu$as using the PASS mode (Riess et al., 2016) and GAIA is expected to achieve 10$\mu$as for bright stars (Bruijne, 2014). However, to achieve sub-$\mu$as astrometric precision needed for earth-like planet detection an absolute reference of the optical distortion is needed to calibrate multi-epoch data.

To overcome this limitation, a concept has been proposed (Guyon et al. 2012a) and demonstrated in the laboratory as part of APRA and TDEM funded efforts (Bendek et al. 2013, and 2017). This approach uses a Diffractive Pupil (DP) to generate precise fiducial features in the image plane, which appear as radial streaks or spikes. These diffractive features can calibrate dynamic or relative distortions since they are imaged by the same optical system, thus serving as a reference for calibration.

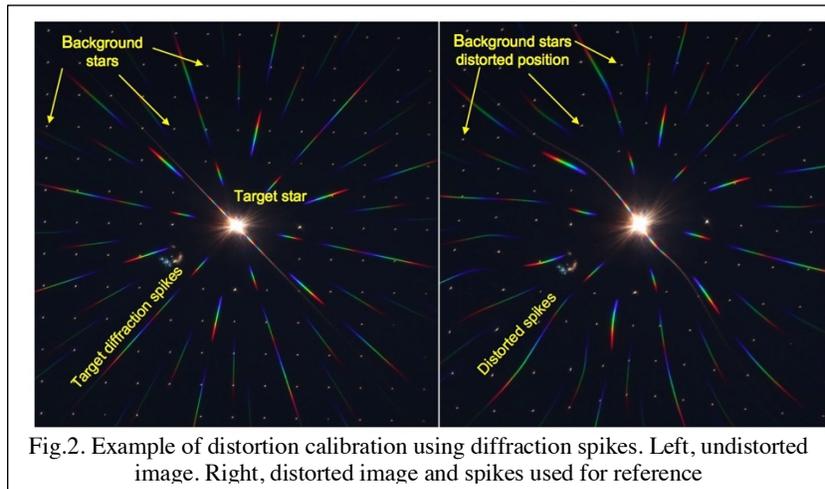

Fig.2. Example of distortion calibration using diffraction spikes. Left, undistorted image. Right, distorted image and spikes used for reference

A TDEM grant has allowed the team to demonstrate the DP technology to medium fidelity. The average accuracy recovered was 5.7x10e-5 λ/D, which is equivalent to 2.5$\mu$as on a 2.4m telescope, or 1.5$\mu$as for a 4m telescope observing in the visible. These results show that distortion calibration can enable future exoplanet mission to detect and measure masses of Earth-like planets around nearby stars. The team has also shown that the DP technology is technically fully compatible, and scientifically synergistic with a coronagraph by performing simultaneous high contrast imaging around the same source. Partial calibration of the optical train can be achieved placing the DP on a reimaged pupil to avoid placing the dots on the primary mirror surface. Trades regarding the placing of the DP and desired accuracy needs to be performed for each mission architectures and scientific goals. Steps to achieve a high-fidelity, sub-$\mu$as equivalent performance have been identified concluding that the most important one is to perform detector metrology and calibration.

*Detector metrology* is needed to achieve sub-$\mu$as Astrometry because CCD/CMOS arrays measure the centroid of stars relative to the position of the pixels in the focal plane. At the $\mu$as level, we can no longer assume the pixels are on a regular rectilinear grid. Typical CCDs have pixel position errors at the ~1% of a pixel. A meter class telescope with a Nyquist sampled focal plane implies centroiding to ~2e-5 pixels. One approach to reducing this error, used on HST is to scan the star images across multiple pixels. This averages out pixel position errors that are "random." It, however, does not address large-scale systematic errors in pixel position over many 1000's of pixel, and it doesn't address the geometry of a mosaic array of detectors.

It is possible to use laser metrology to calibrate, and even monitor in real time (on orbit), the positions of all pixels. The light from two single-mode fibers will produce a set of fringes on a detector. The fringes are moved across the detector, and the phase of the fringes are linearly related to the position of the pixel. A team at JPL led by Dr. Shao has demonstrated repeatability of pixel position measurements to 1e-5 pixels.

Using the DP to correct optical distortions and laser metrology to calibrate the detector it is possible to achieve sub-$\mu$as stellar imaging astrometry precision. However, demonstrating both technologies on the same instrument in a relevant environment still has to be completed during the next decade.

**4. Missions overview**

During this decade a large number of astrometry missions have been proposed by several countries to address a wide range of science cases, including exoplanet detection. The European GAIA is the only astrometry-dedicated mission currently in operation. It provides unprecedented positional and radial velocity measurements of about one billion stars in our Galaxy. The end-of-mission astrometry performance is expected to be about 10μas for stars of G magnitude 6 to 12, and in the order of 30μas for magnitude 15 (Bruijne, 2014). The team expects that 21,000 ± 6,000 high-mass (1-15 Jupiter mass) long-period planets should be discovered out to distances of about 500pc from the Sun (Perryman et al. 2014). However, GAIA will only access Jupiter-size planets. Hence, it will not address the discovery and mass measurement of Earth-like planets.

There are several other astrometry space missions being proposed or developed around the world. The European Space Agency is evaluating the Theia space mission concept (successor of the earlier NEAT concept), primarily designed to study the local dark matter properties and identify Earth-like exoplanets in our nearest star systems.

The Chinese Strategic Pioneer Program (SPP) on Space Science is developing the *Search for Terrestrial Exo-Planet (STEP)* mission, which aims to achieve 1$\mu$as precision enabling earth-like planet discoveries around the nearest stars (Chen, D., 2014).

The Japanese astronomical community is working on the *Japan Astrometry Satellite Mission for Infrared Exploration (JASMINE)* project, which consists of a series of increasingly more capable missions, called Nano, Small and Medium JASMINE (Gouda, N., 2012). Other countries, including Australia and Sweden, are pursuing small mission focused on Alpha Centauri and nearby binary system. Toliman a 30cm aperture telescope for binary stars relative astrometry, is a mission focused on finding Earth-mass planets in the Habitable Zones of the nearest sun-like binary stars. STARE has a very similar approach, but it features a 12.5cm telescope (Janson, M., 2017).

It is worth mentioning that there is no U.S. led astrometry mission, while multiple countries around the world are developing astrometry missions with slight different focus, but all of them have exoplanet science as part of their scientific goals.

Table 1. List of astrometry missions

| Mission | Science goal | Telescope size | technique | status | Country |
|---|---|---|---|---|---|
| GAIA | General astrophysics / Exoplanets | 0.35x0.7m, FoV: scan | Scanning imaging astrometry | In operation | EU |
| Theia | General astrophysics / Exoplanets | 0.8m, 0.5˚FoV | Wide field imaging + laser metrology | Proposed to ESA E5 | EU |
| STEP | Exoplanets around Nearby stars | 1.2m 0.4˚FoV | Wide field imaging + laser metrology | Proposed | China |
| JASMINE | General astrophysics / Exoplanets | 0.05m and 0.3m | Wide field imaging + star cluster calibration | Built and proposed | Japan |
| TOLIMAN | Exoplanets around Nearby stars | 0.3m, ~1' FoV | Binary narrow field astrometry + DP | Proposed | BT/Australia |
| STARE | Exoplanets around Nearby stars | 0.15m ~1' FoV | Binary narrow field astrometry + DP | Proposed | Sweden |

## 5. Conclusion

We have identified the need to measure exoplanet masses to answer important scientific questions such discerning rocky planets from gaseous counterparts and assessing exoplanet habitability. Astrometry is the only technique that can unequivocally measure exo-planet masses, hence, there is a need to advance this technology during the next decade to enable future exoplanet flagship missions such as HABEX and LUVOIR to achieve their full potential. Although several countries have recognized the scientific opportunity and are engaged with various missions which exploit astrometric technologies, the U.S. does not currently have any advanced mission concept in this domain. Here we call upon the community to develop a mission concept or explore partnerships with other countries.